\documentclass[twocolumn]{aastex7}
\usepackage{bm}
\usepackage{ae,aecompl,amsmath}

\begin{document}

\title{On the formation of retrograde S-type planets in binaries with a polar circumbinary disk}
 
\author[orcid=0000-0002-4489-3491,sname='Chen']{Cheng Chen}
\email[show]{cchen@ubishops.ca}
\affiliation{Department of Physics and Astronomy, Bishop's University, 2600 Rue College, Sherbrooke, QC J1M 1Z7, Canada}

\author[orcid=0000-0001-8974-0758,sname='Armitage']{Philip J. Armitage}
\email[hide]{philip.armitage@stonybrook.edu}
\affiliation{Center for Computational Astrophysics, Flatiron Institute, New York, NY 10010, USA}
\affiliation{Department of Physics and Astronomy, Stony Brook University, Stony Brook, NY 11794, USA}

\author[orcid=0000-0002-2137-4146,sname='Nixon']{C. J. Nixon}
\email[hide]{C.J.Nixon@leeds.ac.uk}
\affiliation{School of Physics and Astronomy, University of Leeds, Sir William Henry Bragg Building, Woodhouse Ln., Leeds LS2 9JT, UK}

\author[orcid=0000-0002-4636-7348,sname='Lubow']{Stephen H. Lubow}
\email[hide]{lubow@stsci.edu}
\affiliation{Space Telescope Science Institute, 3700 San Martin Drive, Baltimore, MD 21218, USA}

\author[orcid=0000-0003-2401-7168,sname='Martin']{Rebecca G. Martin}
\email[hide]{rebecca.martin@email.address}
\affiliation{Department of Physics and Astronomy,  University of Nevada, Las Vegas\\ 4505 South Maryland Parkway, Las Vegas, NV 89154, USA}
\affiliation{Nevada Center for Astrophysics, University of Nevada, Las Vegas\\
4505 South Maryland Parkway, Las Vegas, NV 89154, USA}

\begin{abstract}
Retrograde S-type planets have been observed in several binary systems, yet their formation pathway remains poorly understood. With high-resolution hydrodynamic simulations, we demonstrate that a polar circumbinary disk around an eccentric, unequal-mass binary can form and sustain a retrograde mini disk around the primary star. This provides a direct in-situ formation channel for retrograde S-type planets. The mini disk forms via a sub-Keplerian accretion stream that is slightly misaligned from the polar disk. The mini disk initially undergoes von Zeipel–Kozai–Lidov (ZKL) oscillations, driving coupled eccentricity and inclination evolution. 
Rather than oscillating indefinitely,  the inner mini disk evolves past the critical ZKL inclination, decouples from the outer disk, and settles into a stable retrograde orbit. This evolution is sensitive to numerical resolution: the retrograde configuration is absent in previous lower-resolution simulations, where the mini disk accretion timescale is too short to sustain ZKL-driven evolution. For a higher disk viscosity, the mini disk remains near-polar due to a shorter accretion timescale. Since protoplanetary disks typically have low viscosity, our results suggest that retrograde S-type planets can form in-situ from retrograde mini disks around polar circumbinary disks, and their occurrence rate may be higher than currently estimated.
\end{abstract}


\section{Introduction}
Multi-star systems represent a fundamental feature of the universe. Observations indicate that the vast majority of massive stars, along with approximately half of all solar-type stars, belong to binary or higher-order systems \citep[e.g.][]{Raghavan2010, Sana2012}. Planet formation also occurs in these binary systems. Among the 6,138 confirmed planets, approximately 760 reside within binary or higher-order stellar systems. The vast majority of these planets follow S-type orbits, circling just one of the host stars, whereas only around $30$ are found on P-type orbits, an orbit around the entire binary pair (for a comprehensive overview of planet-hosting binaries, see \citealt{Thebault2025}).

Young binary systems are frequently surrounded by circumbinary disks (CBDs), serving as nurseries for P-type planets. Meanwhile, if gas flows onto the binary components and forms circumprimary or circumsecondary disks (hereafter referred to as mini disks), it can facilitate the formation of S-type planets. Most confirmed S-type planets are in prograde (coplanar) orbits with respect to the orbit of their stellar hosts, but some observed systems suggest that retrograde configurations relative to the binary orbit are also possible — notable examples include $\nu$ Octantis Ab \citep{Eberle2010, Gozdziewski2013, Ramm2016, Lee2024} and HD 59686 Ab \citep{Ortiz2016, Trifonov2018}. Beyond the prograde/retrograde distinction, CBDs and mini disks can be misaligned with respect to their binary orbital planes, giving rise to misaligned planets in both P-type \citep[e.g.][]{Czekala2019, Kraus2020, Young2023} and S-type \citep[e.g.][]{Jensen2014, Brinch2016} configurations.

When a mini disk is sufficiently misaligned to the binary orbital plane, it becomes susceptible to von Zeipel-Kozai-Lidov (ZKL) oscillations \citep{vonZeipel1910, Kozai1962, Lidov1962}, a secular mechanism that drives coupled oscillations between the disk inclination and its eccentricity \citep[e.g.][]{Martinetal2014, Fu2015,Fu2015b,Lubow2017, Zanazzi2017, Smallwood2023}. Planet formation under these dynamical conditions can be challenging, as S-type planets assembled from such a dynamically evolving disk inherit this orbital complexity \citep[e.g.][]{Martin2022,Huang2022}, rendering them considerably more vulnerable to long-term orbital instability compared to their coplanar or retrograde counterparts. Prior work shows that P-type planets can be stable at arbitrary inclinations when their semi-major axes are greater than 5 times the binary separation \citep{Doolin2011, Chen20201}. Conversely, S-type planets can be destabilized if their inclination is greater than the critical angle for the ZKL oscillation of 39.2$^{\circ}$ for a prograde orbit or less than 140$^{\circ}$ for a retrograde orbit.

Misaligned CBDs are anticipated to be more prevalent in eccentric binary systems, which characteristically exhibit longer orbital periods \citep{MartinandLubow2017}. In such systems, a polar-aligned configuration---where the disk aligns with the binary's eccentricity vector \citep{Aly2015}---emerges as a third stable state alongside the traditional aligned \citep{Nixonetal2011b} and counter-aligned \citep{Nixonetal2011a} orientations. Due to viscous dissipation, a misaligned CBD can evolve toward a polar-aligned configuration within several hundred binary orbital periods \citep{Aly2015, MartinandLubow2017, Zanazzi:2018}. Consequently, the presence of eccentricity introduces additional dynamical complexities.

SPH simulations in \citet{Chen2026a} have investigated the disk–binary interaction resulting from a polar aligned disc, revealing that the binary's orbit gradually shrinks over time across various binary eccentricities. Additionally, the shrinkage rate for a system with a polar CBD can be faster than for those with prograde or retrograde CBDs. Moreover, model D in \citet{Chen2026a} resolved the formation of a mini disk in an eccentric, unequal-mass binary; this mini disk has a near-polar orbit with respect to the binary orbital plane, but it does not undergo ZKL oscillations. In that simulation, $10^6$ particles were initially used, and the resulting mini disk is composed of only 7,000 particles. To better understand the evolution of this mini disk, higher resolution is required.

In this Letter, we revisit the formation and evolution of mini-disks formed from a polar circumbinary disk with five times more particles compared to previous simulations. This reveals new dynamical behavior that was absent in lower-resolution simulations. In particular, we find that the mini disk initially undergoes ZKL oscillations, driving significant eccentricity and inclination evolution and eventually a retrograde mini disk. We further investigate the role of disk viscosity by varying the \cite{SS1973} $\alpha_{\rm SS}$ parameter, demonstrating its impact on the mini disk structure and the coupling between the inner and outer parts of the mini disk. These results highlight that the long-term evolution of mini disks in eccentric, unequal-mass binaries is sensitive to both numerical resolution and disk physics, with important implications for disk–binary interactions. The organization of this paper is as follows. Section~\ref{sta} outlines the SPH simulation framework of the unequal-mass binary with a polar CBD. Our simulation setup is detailed in Section~\ref{sim}, while Sections~\ref{dis} and \ref{con} provide a discussion and our conclusions, respectively.

\section{simulation setup of an unequal mass binary with a polar CBD}
\label{sta}

This study utilizes the 3D SPH code \textsc{phantom} \citep{Price2018} to model the hydrodynamics of the system. \textsc{phantom} has an established history of application in circumbinary disk studies \citep[starting with][]{Nixon2012}. We initialize a binary system characterized by a total mass $M_{\rm b} = M_1+M_2$, an initial semi-major axis $a_{\rm b}$, an orbital eccentricity $e_{\rm b} = 0.4$, and a mass ratio $q_{\rm b} = M_2/M_1 = 1/3$ for all simulations. A circumbinary disk is initialized on a circular orbit perpendicular to the binary orbital plane. We set the disk mass to a negligibly small value ($M_{\rm d}=10^{-6} M_{\rm b}$) to ensure that the stationary inclination angle remains fixed at approximately 90° relative to the binary orbit.\footnote{Using a low disk mass allows the results to be rescaled. Conversely, massive disks would introduce non-linear effects, such as self-gravity and shifts in the stationary inclination angle, which are not covered here \citep[see e.g.][]{MartinandLubow2019,Abod2022}.}

The disk is initialized within the radial range from $R_{\rm in} = 3~a_{\rm b}$ to $R_{\rm out} = 10~a_{\rm b}$. To model the gas, we adopt an isothermal equation of state, with a sound speed, $c_{\rm s} = 0.1 \sqrt{GM_{\rm b}/R_{\rm in}}$, resulting in a disk aspect ratio of $H/R \approx 0.1$ at the inner edge, $R_{\rm in}$, of the circumbinary disc. As the sound speed is a constant, the circumbinary disk scale height follows the relation $H/R \propto R^{1/2}$. The scale heights of the mini discs that form around each binary component are smaller than the circumbinary disk due to this radial scaling, with a slightly larger normalization due to the lower mass of the object they are orbiting. We conduct simulations at two different resolutions: a low-resolution run with an initial particle number of $10^6$ and a high-resolution run with an initial particle number of 5$\times10^{6}$ SPH particles.

To manage the simulation domain, we enforce an outer boundary at $R_{\rm out}$, where any particles crossing this radius are removed.  The initial surface density profile, $\Sigma(R)$, is constructed following \citet{Nixon2021} \citep[see also the Appendix in][]{Drewes2021}:
\begin{equation}
\resizebox{0.47\textwidth}{!}{$
\Sigma (R)=\left\{
\begin{aligned}
& \frac{\dot{M}_{\rm add}}{3\pi \nu (R)} \left[ 1 - \left(\frac{R_{\rm in}}{R} \right)^{1/2}\right]\frac{R_{\rm out}^{1/2}-R_{\rm add}^{1/2}}{R_{\rm out}^{1/2}-R_{\rm in}^{1/2}} &\ {\rm for} \ R\ \leq \ R_{\rm add} \\
& \frac{\dot{M}_{\rm add}}{3\pi \nu (R)} \left[ \left(\frac{R_{\rm out}}{R} \right)^{1/2} -1 \right]\frac{R_{\rm add}^{1/2}-R_{\rm in}^{1/2}}{R_{\rm out}^{1/2}-R_{\rm in}^{1/2}} &\ {\rm for} \ R\ > \ R_{\rm add},
\end{aligned}
\right.
$}
\end{equation}
where mass is continuously injected into the disk at a radius of $R_{\rm add} = 7a_{\rm b}$ with a rate $\dot{M}_{\rm add}$ and $\nu = \alpha_{\rm SS} c_{\rm s} H$, where $\nu$ is the kinematic viscosity. This is done via the method outlined in the Appendix of \citet{Drewes2021}, but here we set the initial velocity of the injected particles to be the velocity of their nearest neighboring particles, rather than the local Keplerian velocity.

The sink particles representing the binary components are both initially assigned an accretion radius of $R_{\rm acc} = 0.4\ a_{\rm b}$. This relatively large radius is chosen to accelerate the system's convergence to a steady state. We run the simulation with this accretion radius for a time of $1500\,T_{\rm b}$, where $T_{\rm b}$ is the orbital period of the binary. In our simulations, we turn off the back-reaction of the disk on the binary in order to prevent orbital evolution of the binary, allowing the CBD to reach a steady state with the chosen $a_{\rm b}$.

Once the CBD has reached a steady state, we reduce the accretion radii to investigate the detailed gas dynamics within the vicinity of the binary components. We reduce $R_{\rm acc}$ to a fraction of the respective Roche lobe radii, $R_{\rm acc} = 0.1\,R_{\rm RL}$, where $R_{\rm RL}$ is defined by \citep{Eggleton1983}
\begin{equation} \label{egg}
R_{\rm RL} = \frac{0.49\, q_{\rm b}^{2/3}}{0.6\, q_{\rm b}^{2/3}+\ln({1+q_{\rm b}^{1/3}})} a_{\rm b}.
\end{equation}
Therefore we take, $R_{\rm acc} = 0.0476\ a_{\rm b}$ and $0.0289\ a_{\rm b}$ for the primary and secondary, respectively (with $q_{\rm b} \to 1/q_{\rm b}$ is used in eq.~\ref{egg} for the primary). We define $t = 0$ as the moment when $R_{\rm acc}$ is reduced. 

We model physical viscosity using the Navier-Stokes equation, parameterized by the Shakura-Sunyaev viscosity, $\alpha_{\rm SS}$ \citep{SS1973}. Our study explores values of $\alpha_{\rm SS}=0.1$ and $0.3$. Regarding the artificial viscosity in SPH simulations, we employ both linear and quadratic terms. The linear coefficient for each particle, $\alpha_{\rm SPH}$, is regulated by a modified time-dependent Cullen-Dehnen switch \citep{Cullen2010, Price2018}, confined within the range $[0.01, 1.0]$, while the quadratic term is linked to the linear term via $\beta_{\rm SPH}=2\alpha_{\rm SPH}$. This adaptive viscosity scheme effectively minimizes dissipation in smooth flows while providing necessary shock capturing \citep[e.g.,][]{Chen2025a}.

\section{Simulation results}
\label{sim}
This section presents the outcomes of our SPH simulations. We collect our discussion of the results into two subsections corresponding to two different values of $\alpha_{\rm SS}$: the fiducial model with $\alpha_{\rm SS} = 0.1$ (model A) and the higher value $\alpha_{\rm SS} = 0.3$ (model B).

\subsection{Model A: fiducial $\alpha_{\rm SS}$ = 0.1 }

\begin{figure*}
    \centering
    \includegraphics[width=0.48\linewidth]{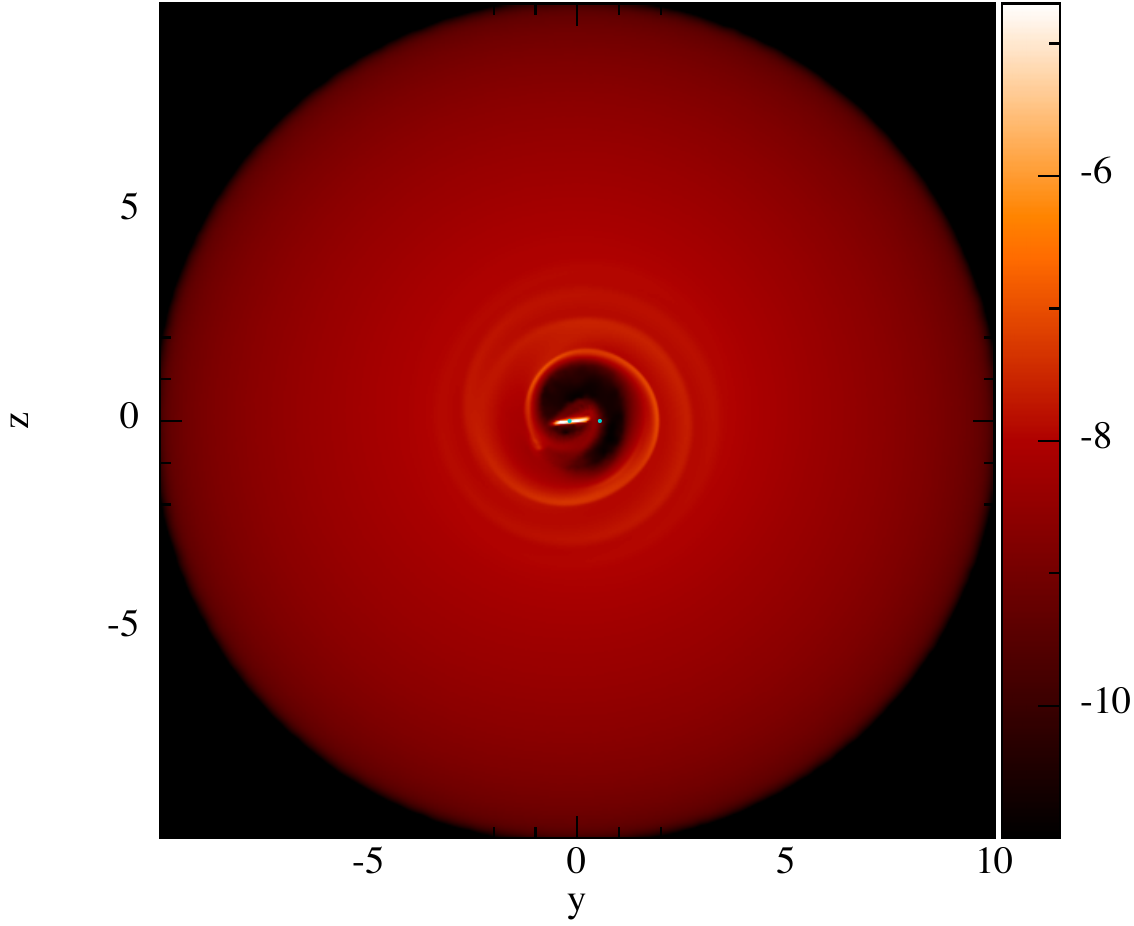}
    \includegraphics[width=0.48\linewidth]{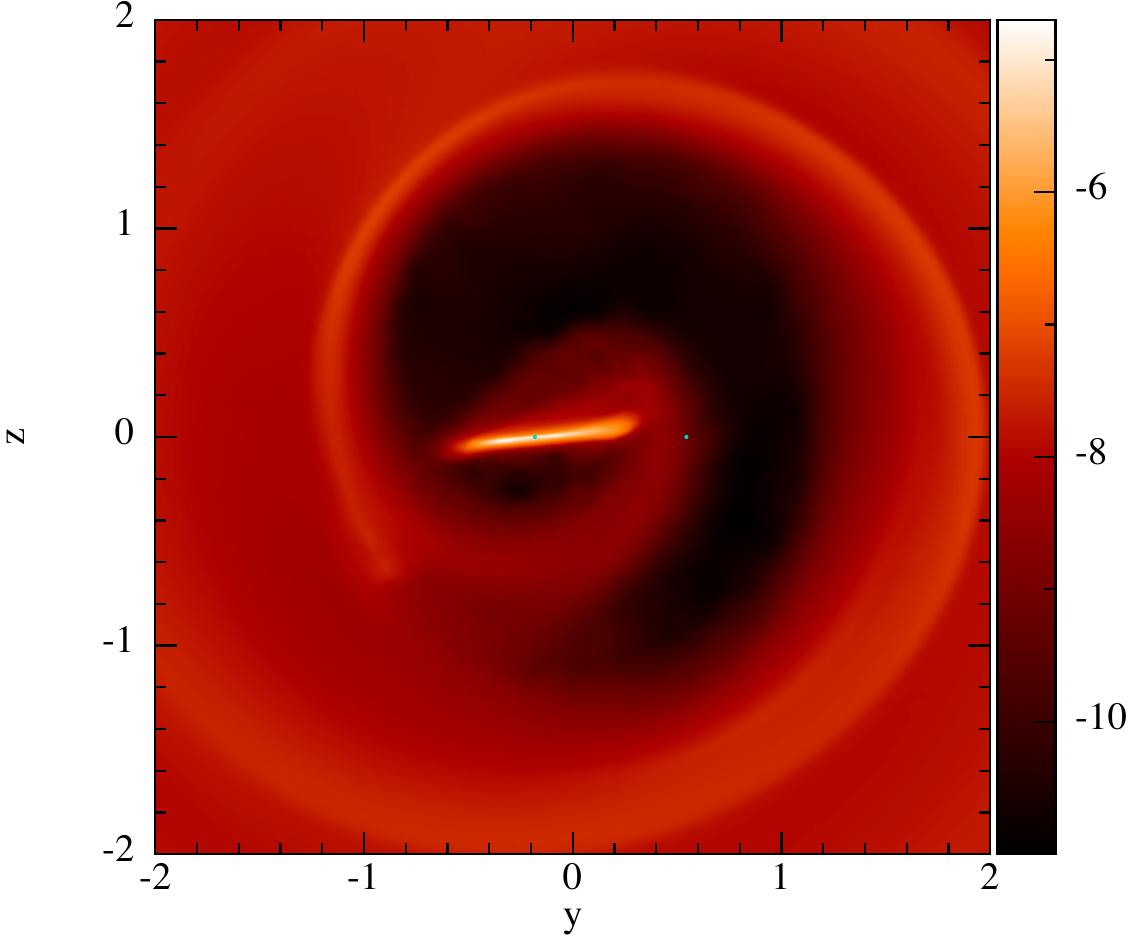}
    \includegraphics[width=0.48\linewidth]{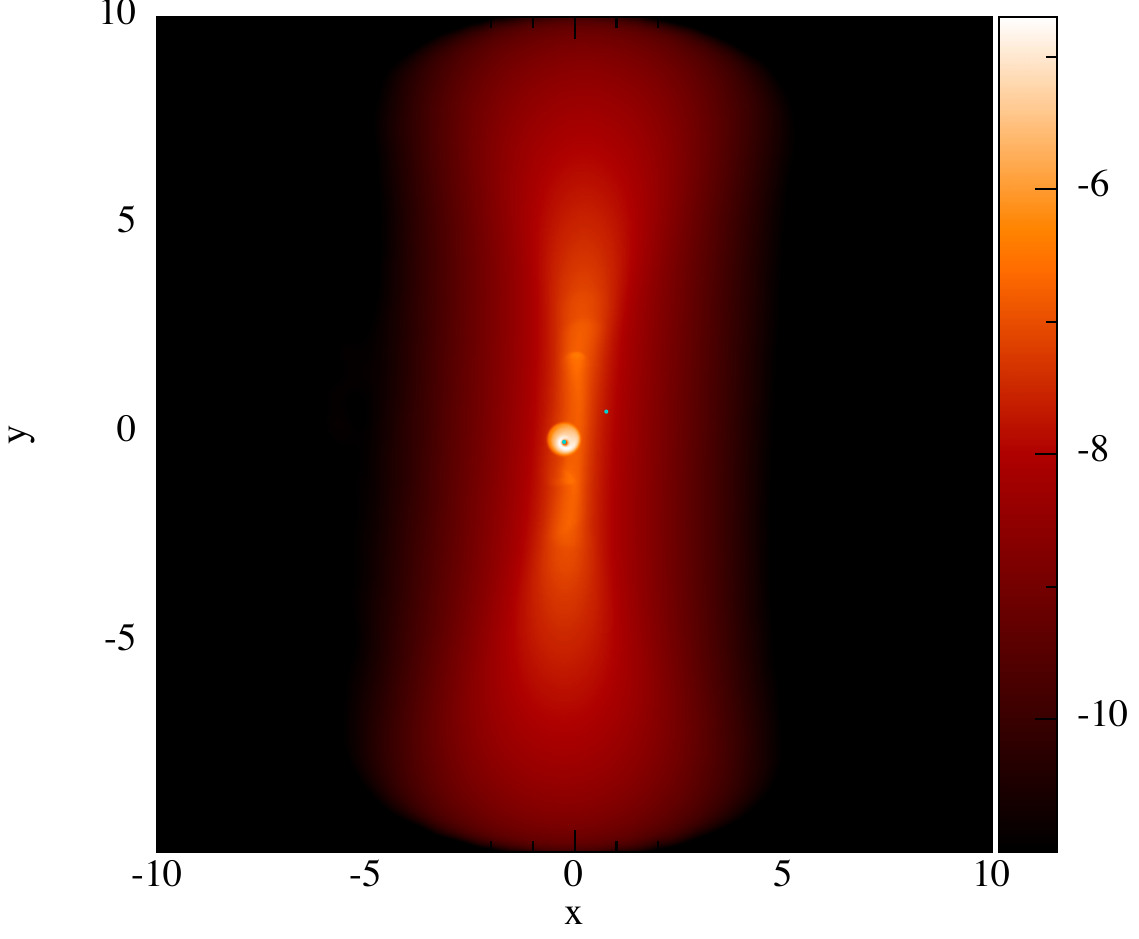}
    \includegraphics[width=0.48\linewidth]{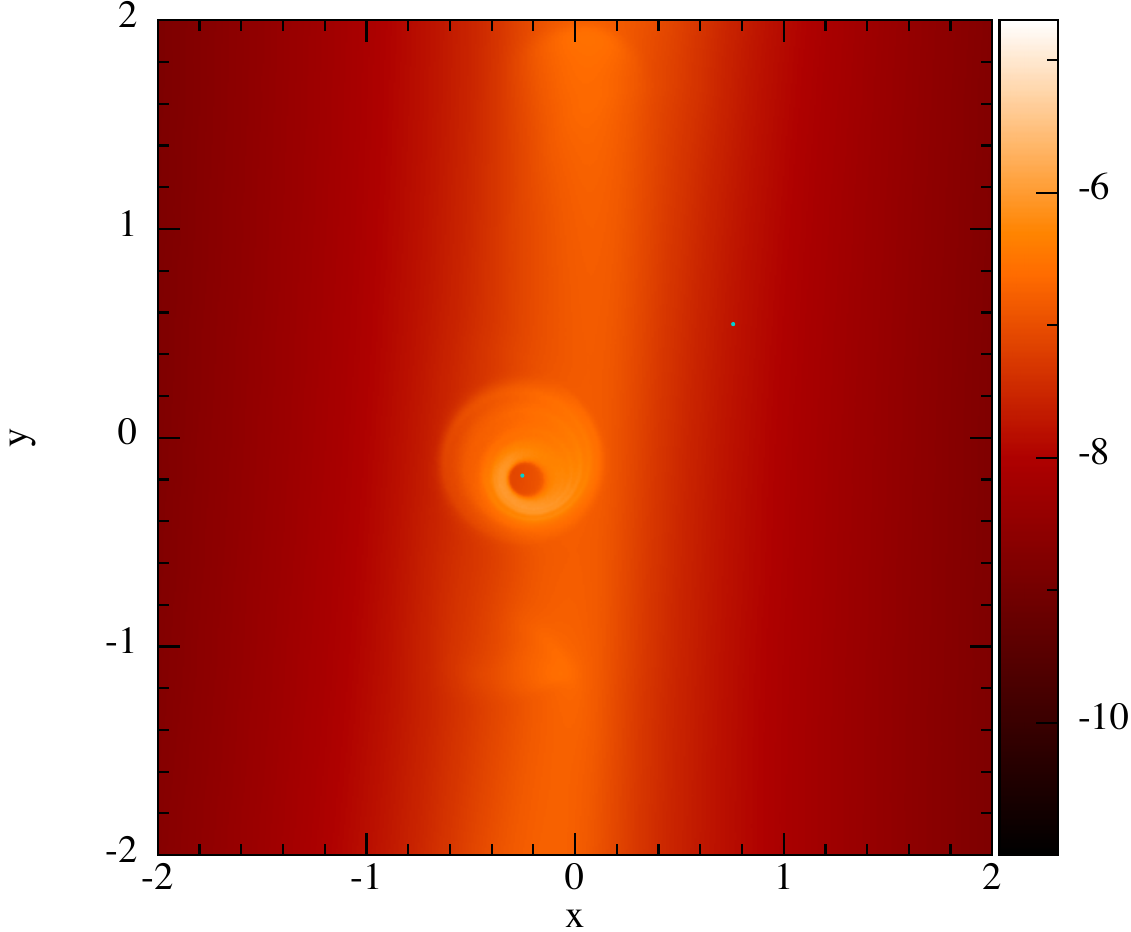}
    \includegraphics[width=0.48\linewidth]{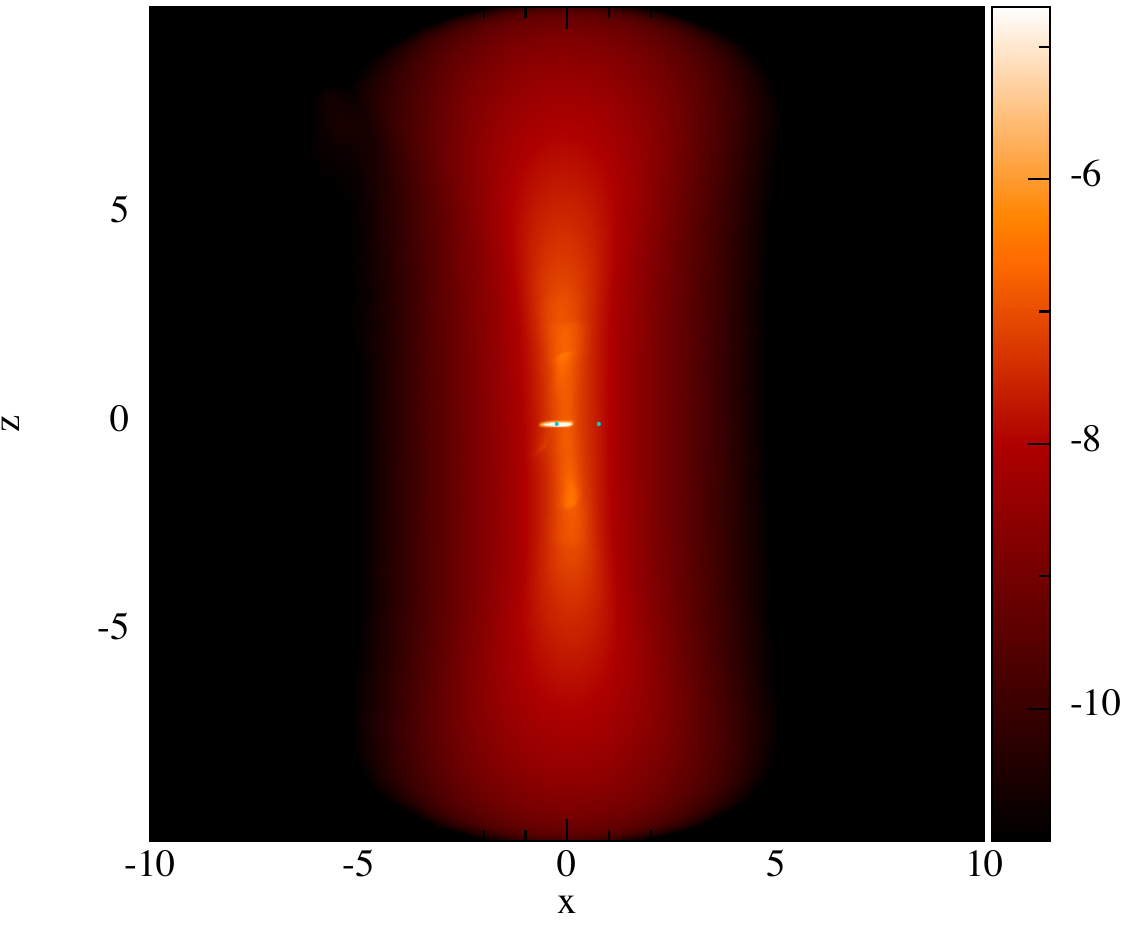}
    \includegraphics[width=0.48\linewidth]{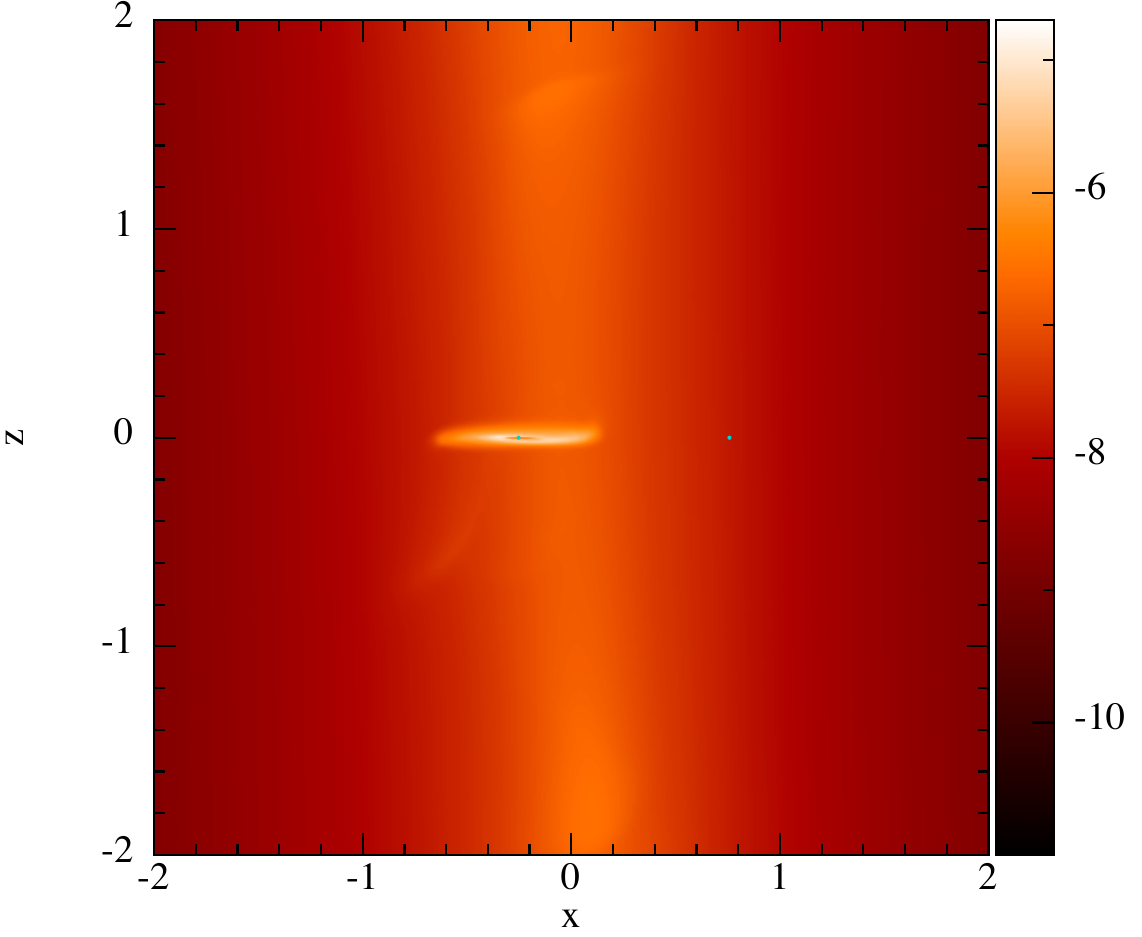}

    \caption{Column density plots for the simulation with $\alpha_{\rm SS}=0.1$ 
    in the y-z plane (upper), the x-y plane (middle) and the x-z plane (lower) at time $t=100\ T_{\rm b}$. The right panels are the zoom-in plots within $\pm$ 2 $a_{\rm b}$. The two cyan circles represent the location of the binary stars (sink particles) with sizes equal to the size of the accretion radius of each star. The unit of the axis is $a_{\rm b}$ and the color bars are the same for all of the panels with the density in arbitrary units.}
    \label{fig:A}
\end{figure*}

We show the disk simulation with $\alpha_{\rm SS}=0.1$ in Fig.~\ref{fig:A} at a time $t=100\ T_{\rm b}$. After we reduce $R_{\rm acc}$, the number of particles in the disk $N_{\rm p}$ increases from  $5.0\times 10^{6}$ to $7.4\times 10^{6}$. 
We define a Cartesian coordinate system $(x, y, z)$ centered on the 
center of mass of the binary, with the $x$-$y$ plane coinciding with the binary orbital plane and the $z$-axis aligned with the binary  orbital angular momentum vector.  The binary is initialized at apoapsis, with the primary located 
at $(-0.35\,a_{\rm b},\, 0,\, 0$) and the secondary at $(+1.05\,a_{\rm b},\, 0,\, 0)$. Fig.~\ref{fig:A} shows the column density of the simulation in the $x$-$z$, $x$-$y$, and $y$-$z$ planes from upper to lower panels, respectively, while the right panels are zoom-in views of the left panels. 
The cavity around the binary has a size of $R\sim2\,a_{\rm b}$ and a stream flows from the inner edge of the CBD to the primary star of the binary. A mini disk then forms around the primary via this stream after we reduce $R_{\rm acc}$. 

\begin{figure}
    \centering
    \includegraphics[width=1.0\linewidth]{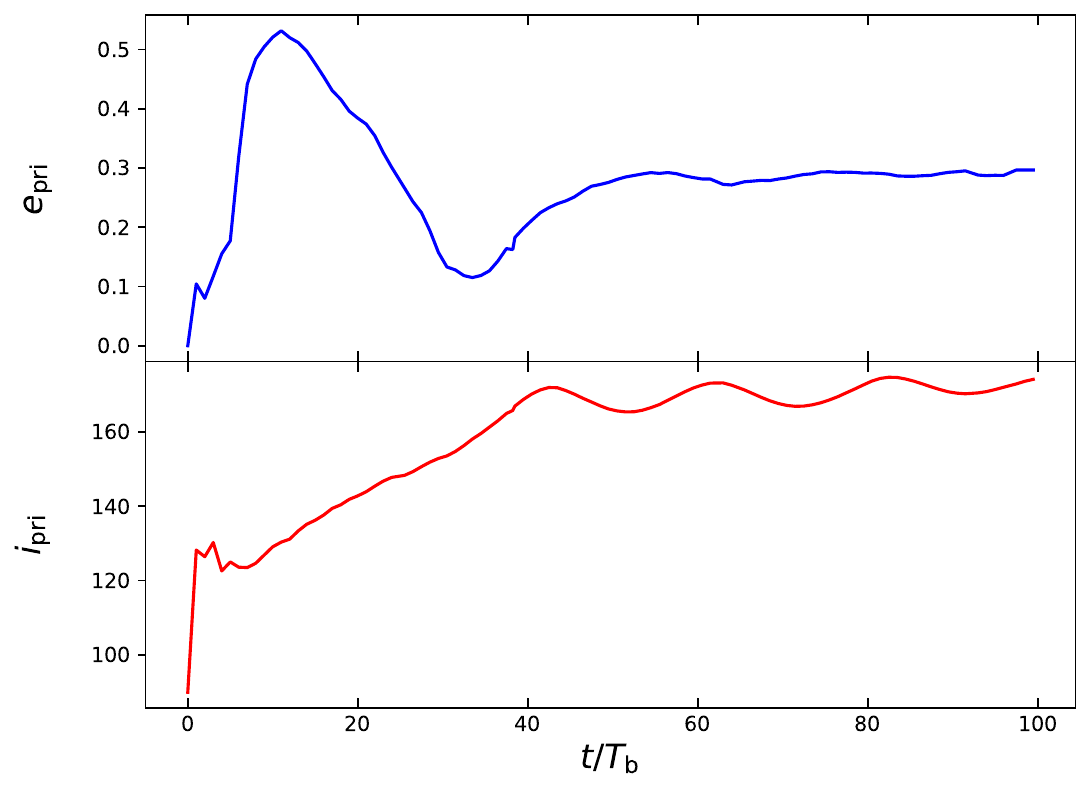}
    \caption{Time evolution plots for the eccentricity $e_{\rm pri}$ (upper) and inclination $i_{\rm pri}$ (lower) of the circumprimary mini disk  at a radius of $0.3\,R_{\rm RL}$ for model A with the smaller $R_{\rm acc}$. After a time of 50$\,T_{\rm b}$, it settles into a configuration with $e_{\rm pri} \approx 0.2-0.25$ and $i_{\rm pri} \approx 170^\circ$.}
    \label{fig:Aevo}
\end{figure}

Fig.~\ref{fig:Aevo} shows the inclination and the eccentricity of the mini disk around the primary, measured at a radius of $R_{\rm pri} = 0.3\,R_{\rm RL}$. The inclination (lower panel) evolves from an initially polar orbit toward a nearly retrograde orbit, from $90^{\circ}$ to $170^{\circ}$, while the eccentricity of the mini disk (upper panel), $e_{\rm pri}$, rises from 0 to above 0.5 and then drops to about 0.2. After a time of $30\,T_{\rm b}$, $i_{\rm pri}$ oscillates around $170^{\circ}$ and the mini disk is composed of approximately 320,000 particles. Our simulation does not resolve the mini disk around the secondary, owing to its smaller Roche lobe radius which cannot support sufficient long-lived material within the binary orbit.

In order to better understand the formation and evolution of the primary mini disk, we compute the specific angular momentum of each SPH particle as $\ell = |\mathbf{r} \times \mathbf{v}|$, where $\mathbf{r}$ and $\mathbf{v}$ are measured relative to the center of mass of the binary. We normalize this by the local Keplerian specific angular momentum,
\begin{equation}
\ell_{\rm kep}(r) = \sqrt{G M_{\rm b} r}.
\end{equation}
The ratio $\ell / \ell_{\rm kep}$ thus quantifies the degree to which a particle's angular momentum deviates from a locally circular Keplerian orbit.

\begin{figure*}
    \centering
    \includegraphics[width=1.0\linewidth]{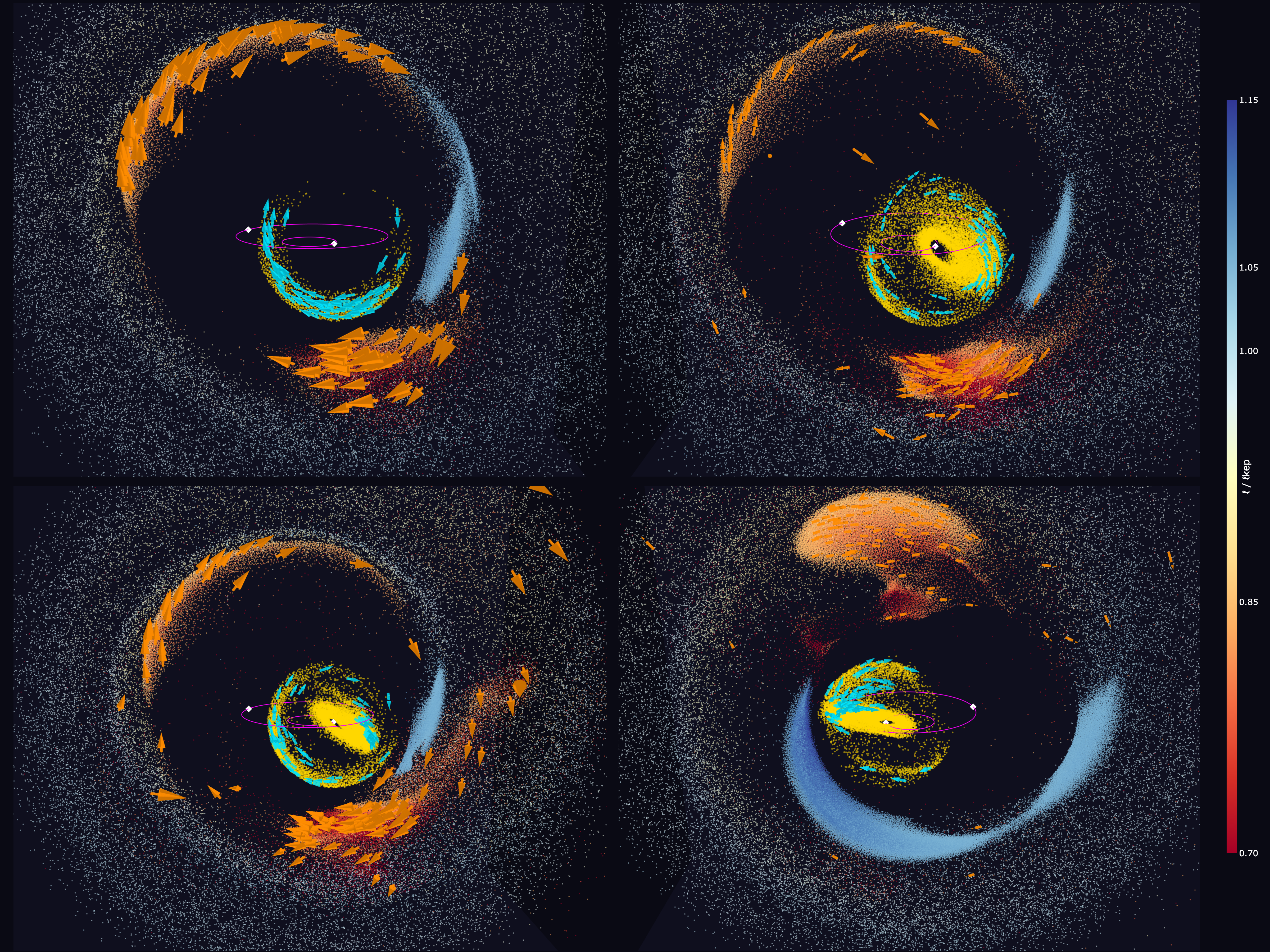}
    \caption{Snapshots of the innermost region ($R \leq 3\,a_{\rm b}$) of the simulation with $\alpha_{\rm SS}=0.1$
    at $t = 0\,T_{\rm b}$ (upper-left), $10\,T_{\rm b}$ (upper-right), $20\,T_{\rm b}$ (lower-left), and $85\,T_{\rm b}$ (lower-right). The time $t = 0$ is defined as the moment when $R_{\rm acc}$ is reduced to the smaller value. In each panel, the two white dots represent the locations of the primary and secondary stars. The bright magenta curves in each panel show the orbits of the two binary components (x--y plane).}    
    Yellow dots indicate particles in the mini disks, while the colors of other particles represent their relative specific angular momentum $\ell / \ell_{\rm kep}$. Cyan and orange arrows denote the velocity directions of the mini disk and the CBD, plotted only for particles at the outer edge of the mini disks and those with $\ell / \ell_{\rm kep} < 0.85$.    
    \label{fig:time}
\end{figure*}

Fig.~\ref{fig:time} shows four snapshots of the innermost region of our simulation ($R\leq3\,a_{\rm b}$) at $t = 0$ (upper-left), 10 (upper-right), 20 (lower-left), and 85 (lower-right) $T_{\rm b}$. In each panel, two white dots represent the locations of the primary and secondary stars. The bright magenta curves in each panel show the orbits of the two binary components; both orbits lie within the binary orbital (x-y) plane, and the vertical direction relative to the curves corresponds to the z-axis. Note that the x-y plane does not coincide with the horizontal plane shown in each panel; the magenta curves themselves indicate the true orientation of the binary orbital plane in each view. Yellow dots indicate particles in the mini disk ($R_{\rm pri} < 1.5\,R_{\rm RL}$), while the colors of other particles represent $\ell / \ell_{\rm kep}$. Cyan and orange arrows denote the velocity directions, plotted only for particles at the outer edge of the mini disk and those with $\ell / \ell_{\rm kep} < 0.85$.

The upper-left panel shows that a misaligned asymmetric arc ($i_{\rm pri} > 90^{\circ}$) has already formed around the primary before we reduce $R_{\rm acc}$. There are two misaligned sub-Keplerian streams. The upper one has an inclination of about $90^{\circ}$ on average, with the lower one being more misaligned than the upper one by + 15$^{\circ}$ on average. As particles from the CBD approach the binary ($R < 2\,a_{\rm b}$), their orbits become increasingly eccentric. Consequently, a super-Keplerian stream forms due to the gravitational slingshot of the binary. This super-Keplerian stream (blue particles) repeatedly impacts the innermost edge of the CBD, producing misaligned sub-Keplerian streams along the edge of the CBD.

Subsequently, as the super-Keplerian stream collides with the sub-Keplerian stream, the front of the super-Keplerian stream merges into the sub-Keplerian stream, producing a more misaligned sub-Keplerian branch. This misaligned branch forms an asymmetric arc around the primary and eventually accretes onto it (we show the details in Section~\ref{dis}).

At a time of $10\,T_{\rm b}$ (upper-right), the mini disk has formed and its inner part has evolved to a more misaligned orbit than the outer part. At $20\,T_{\rm b}$ (lower-left), the inner mini disk has evolved to near a retrograde orbit and decoupled from the outer ring. At $85\,T_{\rm b}$ (lower-right), most of the mini disk has evolved to a retrograde orbit, with some residual misaligned particles remaining around it.

\subsection{Model B: higher $\alpha_{\rm SS}$ = 0.3}

\begin{figure}
    \centering
    \includegraphics[width=\linewidth]{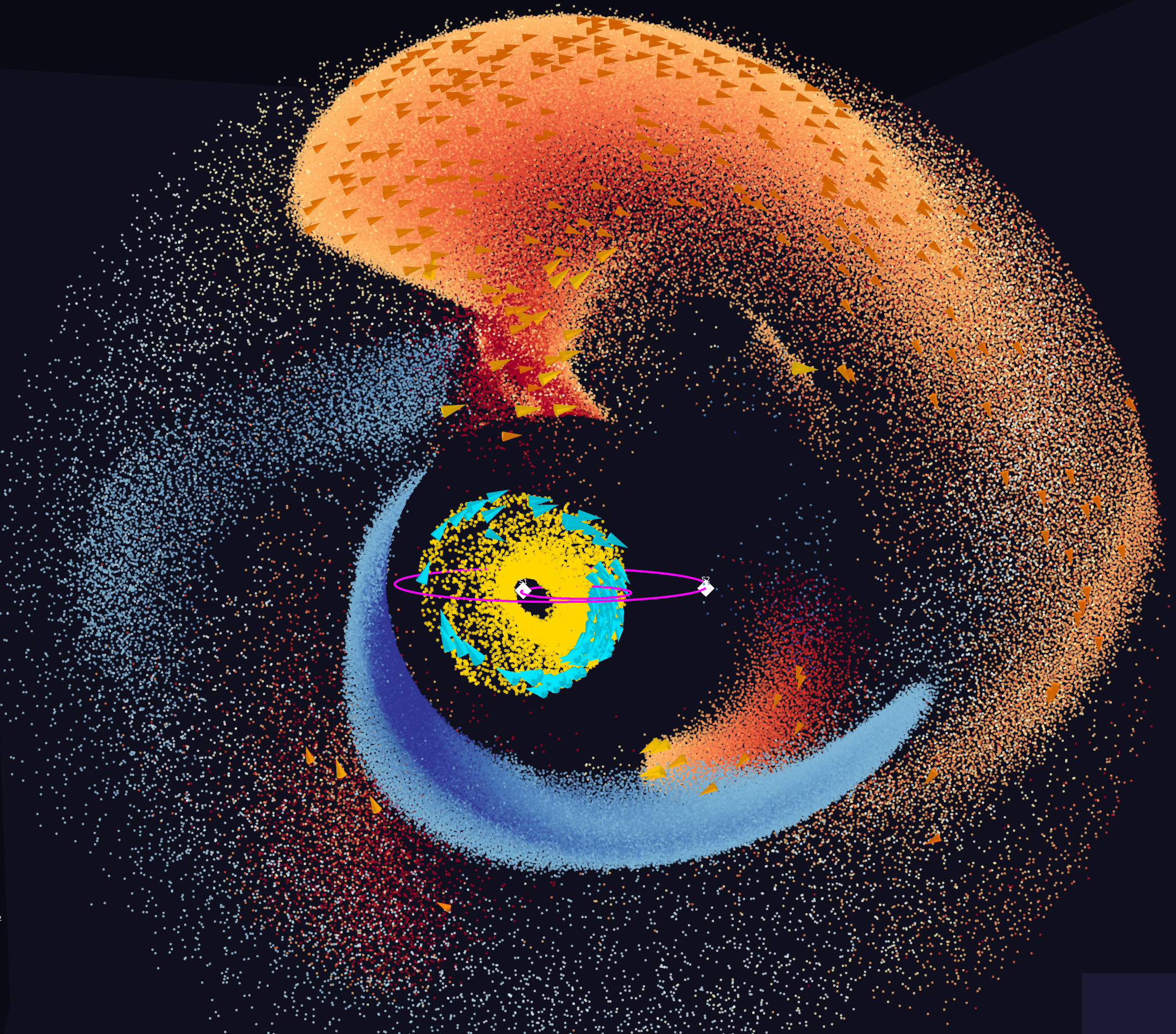}
    \caption{Same as Fig~\ref{fig:time} but for model B with $\alpha_{\rm SS}=0.3$ at a time of $t=50\,T_{\rm b}$. }
    \label{fig:modelB}
\end{figure}

For the disk with a larger $\alpha_{\rm SS} = 0.3$, the higher viscosity drives a larger accretion rate through the circumbinary disk than in Model A, reaching $8.7\times10^{-9}\ m_{\rm b}/T_{\rm b}$, which is four times higher.\footnote{Note that the change in accretion rate is non-linear with respect to the disc viscosity due to the complex nature of binary disc interaction which includes, for example, that the value of the viscosity parameter affects how quickly and where the energy and angular momentum transferred from the binary to the disc is dissipated in the disc \citep[e.g.][]{Heath2020}.} In Fig.~\ref{fig:modelB}, the mini disk forms after reducing $R_{\rm acc}$ to $0.1\ R_{\rm RL}$, with approximately 181,000 particles and is wrapped around by the outer mini disk at $t=50\,T_{\rm b}$. However, the inner and outer parts do not decouple from each other.

\begin{figure}
    \centering
    \includegraphics[width=1.0\linewidth]{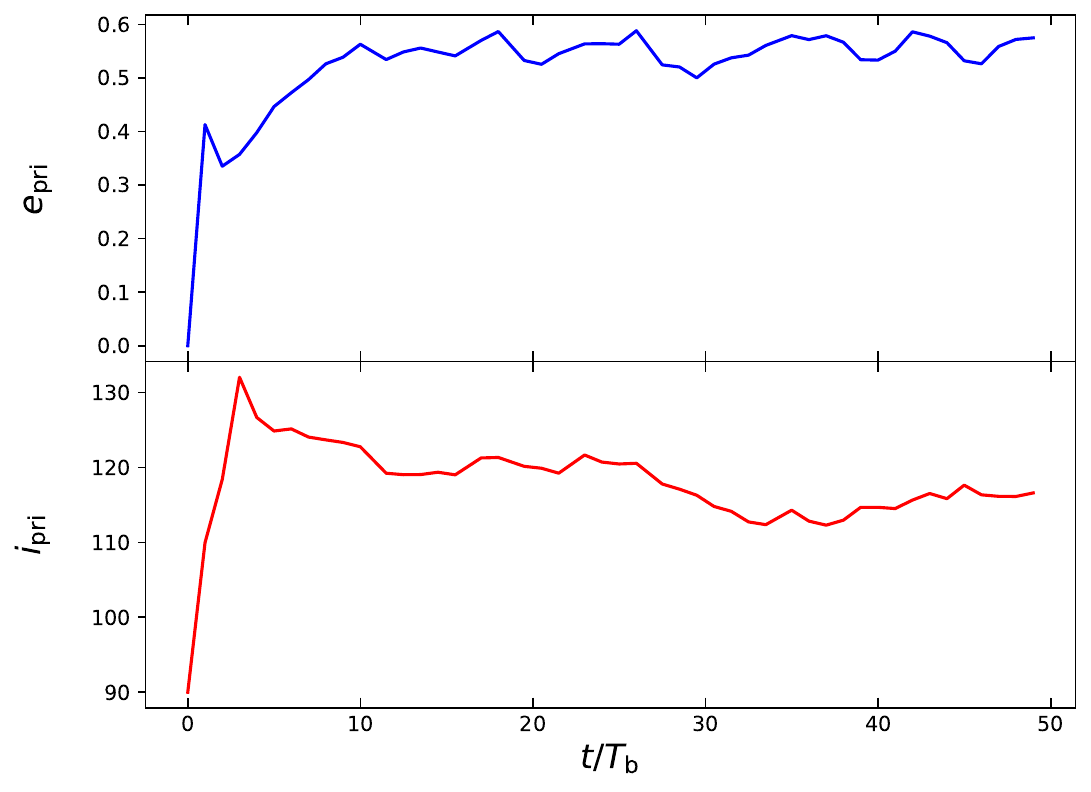}
    \caption{Same as Fig.~\ref{fig:Aevo} but for model B with $\alpha_{\rm SS}=0.3$.}
    \label{fig:Bevo}
\end{figure}

Fig.~\ref{fig:Bevo} shows the eccentricity $e_{\rm pri}$ and inclination $i_{\rm pri}$ of the mini disk at $R_{\rm pri} = 0.3 R_{\rm RL}$. The eccentricity  oscillates around 0.55 and is more eccentric than that of Model A. Initially, $i_{\rm pri}$ increases to $130^{\circ}$ due to the ZKL oscillation and later evolves toward $\sim$$115^{\circ}$.

The accretion timescale $M_{\rm disk}/\dot{M} \approx 4.1\ T_{\rm b}$ is shorter than the ZKL oscillation timescale. Consequently, the mini disk does not fully decouple or evolve to a retrograde orbit, instead remaining in a near-polar, slightly warped configuration. This accretion timescale is close to that in the low resolution simulation with $\alpha_{\rm SS}=0.1$ in \citet{Chen2026a}, implying that the effective viscosity in the lower resolution simulation is comparable to $\alpha_{\rm ss}$ = 0.3 with high resolution.

\section{Discussion}
\label{dis}

\subsection{Retrograde Mini Disk Formation Mechanism}

Compared with the previously presented  simulations in \citet{Chen2026a}, we have increased the number of particles initially in the circumbinary disk $N_{\rm p,ini}$ by a factor of 5. The greatest difference between the low- and high-resolution simulations is that the mini disk around the primary star evolves to a retrograde orbit rather than remaining near a polar orbit (see Figure 6 in \citealt{Chen2026a}). The particle number of the mini disk in the simulation here is approximately 350,000 ($\sim 7\times10^{-8}\,m_{\rm b}$), while the accretion rate is approximately $2.0\times10^{-9}\,m_{\rm b}/T_{\rm b}$. Hence, the accretion timescale\footnote{The mini disk accretion timescale is the ratio between the mini disk mass and the mini disk accretion rate.} is about $35\,T_{\rm b}$, which is approximately 11 times longer than that of the low-resolution case\footnote{The accretion timescale of the mini disk in \citet{Chen2026a} is incorrect because we omitted a factor of $2\pi$ in the denominator. The correct accretion timescale should be $20/2\pi \sim 3.2\,T_{\rm b}$.}.



The initial misalignment of the mini disk is slightly greater than $90^{\circ}$. This is  within the region of retrograde ZKL oscillations \citep{Martin2016}. Hence, the disk exchanges eccentricity and inclination and evolves towards 
the critical ZKL angle of $i_{\rm pri}\approx 140^\circ$.
This is possible because the accretion timescale is greater than the ZKL timescale. 
N-body simulations show that the ZKL oscillation timescale is slightly larger than $20\,T_{\rm b}$ \citep[see Figure~5 in][]{Chen2026a}.
However, rather than sustaining periodic inclination--eccentricity cycles, the inner part of the mini disk decouples from the outer part and undergoes a direct alignment toward the retrograde plane. This rapid evolution is governed by the strong differential precession induced by the binary's secular torque, which generates disk warps that are subsequently damped by internal viscous dissipation. By continuously extracting the system's mechanical energy, this dissipative process ultimately aligns the mini disk into the retrograde coplanar configuration, representing the global energy minimum for the system. 

The ZKL oscillation timescale for an unequal-mass binary is longer than that for the equal mass case \citep{Fu2015b}. Hence, the secular torque from the binary may drive the mini disk inclination past the critical angle for retrograde ZKL oscillations (140$^{\circ}$) before a complete ZKL cycle can develop, facilitating the transition to the retrograde configuration. We note, however, that our equal-mass binary simulation does not resolve the mini disk around the primary star, preventing a direct comparison. Further simulations are required to fully characterize the dependence of the ZKL timescale on binary mass ratio.



Apart from the ZKL dynamics, the long-term sustainability of the retrograde mini disk depends on the continuous replenishment of misaligned material from the CBD. The origin of this misaligned sub-Keplerian stream  can be understood as follows.


Resonant torques from the binary distort the inner edge of the CBD, giving rise to a persistent overdensity, or ``lump'', that orbits at approximately the local Keplerian frequency \citep[e.g.][]{Shi2012, Roedig2012, Miranda2017}. As gas accumulates in the lump, it eventually penetrates the tidal barrier of the binary and flows towards the binary, whereupon gravitational torques impart excess angular momentum to the gas via a slingshot mechanism, launching a super-Keplerian stream radially outward. 


In the current case of a polar disk, the outgoing stream also subsequently re-impacts the inner edge of the CBD, driving strong shocks that dissipate energy and remove angular momentum from the gas. The post-shock material, now sub-Keplerian, loses sufficient angular momentum to partially decouple from the CBD. This sub-Keplerian stream then collides with the super-Keplerian stream again, becoming more misaligned and falling inward as an accretion stream toward the primary. This lump formation process occurs on a timescale much shorter than $T_{\rm b}$. On the other hand, particles in the super-Keplerian stream can be deflected by the binary (or primary at periastron, resulting in changes of the angular momentum in different directions. Hence, we infer this effect causes the misalignment of the stream. Future studies will quantitatively measure the level of this effect in prograde, polar and retrograde CBD.

\begin{figure*}
    \centering
    \includegraphics[width=1.0\linewidth]{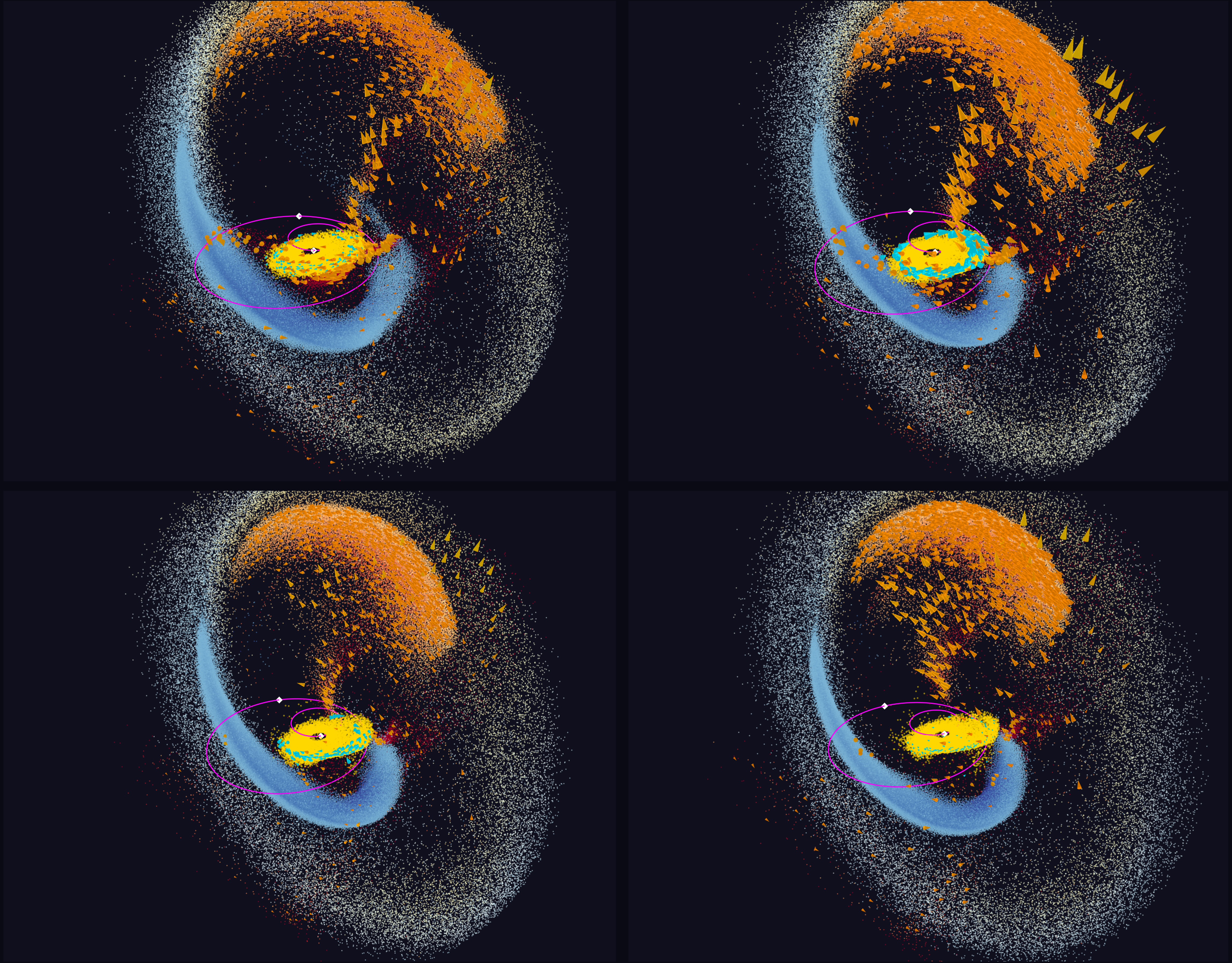}
    \caption{Same as Fig.~\ref{fig:time} but for times $t=86.85$ (upper-left), 86.87 (upper-right), 86.89 (lower-left), and 86.91 $\,T_{\rm b}$ (lower-right).}
    \label{fig:t86}
\end{figure*}

In Fig.~\ref{fig:t86}, the upper-left panel shows the front of the super-Keplerian stream colliding with the sub-Keplerian stream at $t = 86.85\,T_{\rm b}$. In the following panels, the front of the super-Keplerian stream merges into the sub-Keplerian stream, forming a more misaligned branch. This branch subsequently collides with the super-Keplerian stream again and falls toward the primary. Hence, the retrograde mini disk can be sustained over a long time through the continuing replenishment by this misaligned sub-Keplerian branch. It provides the angular momentum necessary to maintain the retrograde configuration.

\subsection{The effect of $\alpha_{\rm ss}$}
We note that although the fiducial viscosity used here ($\alpha_{\rm SS}$ = 0.1) is higher than typical estimates for protostellar disks, we expect the qualitative outcome---that sufficiently low viscosity favors the retrograde-alignment channel over the near-polar outcome---to persist at lower $\alpha_{\rm SS}$. Both the mini disk accretion timescale and the viscous timescale governing (linear) warp damping scale approximately as 1/$\alpha_{\rm SS}$ \citep[e.g.][]{Bate2000}, so their ratio, which determines whether the disk completes its dynamical evolution before being accreted, does not change significantly for lower $\alpha_{\rm SS}$.
Moreover, the alignment mechanism identified here relies on the damping of large-amplitude warps driven by strong differential precession, a regime in which non-linear dissipation channels (such as supersonic shear instabilities, \citealt{Drewes2021}, and parametric instability, \citealt{Gammieetal2000,Ogilvie2013b}) can supplement the linear viscous damping and prevent the alignment timescale from scaling unfavorably as $\alpha_{\rm SS}$ decreases. A dedicated exploration of the $\alpha_{\rm SS}$-dependence at lower viscosity, ideally combined with even higher numerical resolution, is left for future work.

\subsection{Formation of retrograde S-type planets}

Polar-aligned CBDs have been found in some systems, such as HD 98800 \citep{Kennedy2019}, V773 Tau B \citep{Kenworthy2022}, and the polar-aligned debris disk around the binary 99 Herculis \citep{Kennedy2012, Smallwood2020}. Recent observations also reveal significant misalignment between binary orbital planes and disk planes in young stellar systems for binaries with orbital periods greater than 30 days \citep{Czekala2019}. Previous simulations also showed that a misaligned CBD around an eccentric binary could evolve toward the polar configuration within several hundreds binary orbital periods \citep[see][for the details]{Aly2015, MartinandLubow2017}. Recent N--body simulations also found the fraction of polar CBDs compared to coplanar CBDs is sensitive to $e_{\rm b}$ for low mass disks \citep{Johnson2025}. Hence, we expect polar CBDs around binaries to be common. On the other hand, no retrograde CBD has been confirmed in young stellar systems. The DX Cha binary system is one possible candidate that may host a retrograde CBD, as recently reported \citep[see][for details]{Chen2026b}, but more observational evidence is required to verify this.

In principle, a binary system hosting a coplanar CBD should have coplanar mini disks around its binary components, and vice versa. However, a misaligned CBD tends to evolve toward a polar orbit due to viscous dissipation within several hundred $T_{\rm b}$ \citep{MartinandLubow2017}. Such misaligned CBDs typically reside in polar libration orbits \citep[see][for details]{Farago2010, Doolin2011, Chen20192}. In order to form a retrograde CBD, the initial misalignment must exceed $100^{\circ}$ for $e_{\rm b} = 0.2$, and must be even larger for more eccentric binaries. Otherwise, a CBD will evolve toward either a polar or coplanar configuration. Hence, the formation of a retrograde CBD may be less common than prograde or polar.

Recently, observations found some S-type planets may reside in retrograde orbits, such as $\nu$ Octantis Ab \citep{Eberle2010, Gozdziewski2013, Ramm2016} and HD 59686 Ab \citep{Ortiz2016, Trifonov2018}. Additionally, three-body simulations show that a candidate Jupiter-mass S-type planet HD 175370 b could only be stable if it is in a retrograde orbit \citep{Wood2026}. Previous {\sc N}-body simulations suggested that these retrograde S-type planets captured from prograde circumbinary orbits into retrograde circumstellar orbits \citep{Gong2018}, but it requires strict physical conditions. Therefore, the simulations in this study demonstrate that a polar CBD can host a retrograde mini disk in situ, providing a plausible formation pathway for retrograde S-type planets in binary systems. Consequently, more retrograde S-type planets may yet be discovered, and the true abundance of this class of planets could be higher than currently estimated.

\section{Conclusion}
\label{con}

We have investigated mini disks in polar circumbinary disk systems with an eccentric, unequal-mass binary using SPH simulations. By increasing the initial number of particles by a factor of five compared to the previous simulation in \citet{Chen2026a}, a denser mini disk forms around the primary. This mini disk subsequently evolves to a more misaligned orbit while its inner part decouples from the outer part. Eventually, the mini disk evolves to a retrograde orbit with respect to the binary orbital plane. We find that the accretion timescale of the mini disk is approximately 1.75 times its ZKL oscillation timescale. However, rather than undergoing continuing ZKL oscillations, the mini disk evolves into a retrograde orbit that provides a more stable environment for the formation of retrograde S-type planets. We also consider the case of higher $\alpha_{\rm SS} = 0.3$, but the accretion timescale is too short to allow the mini disk to evolve. As a result, the mini disk with the higher $\alpha_{\rm SS}$ remains near a polar orbit. Since protoplanetary disks typically have lower $\alpha_{\rm SS}$, our models imply that retrograde S-type planets can form from retrograde mini disks around polar CBDs.

The upcoming fleet of next-generation space-based and ground-based observatories, such as the Nancy Grace Roman Space Telescope \citep{Spergel2015}, the European Space Agency's PLAnetary Transits and Oscillations of stars (PLATO) mission \citep{Rauer2014}, and the Vera C. Rubin Observatory \citep{Ivezic2019}, are poised to deliver an unprecedented volume of photometric and astrometric data. Moreover, China's Earth 2.0 (ET) mission will be equipped with six ultra-wide-field transit telescopes and two microlensing telescopes \citep{Ge2022}. These missions will not only increase the total number of known exoplanets by tens of thousands but also push the detection limits toward true Earth analogs; consequently, more misaligned and retrograde S-type planets will be discovered in binary systems. We expect more retrograde S-type planets to be found, as polar CBDs are common in the Universe and provide a new possible formation pathway for this class of planets. These retrograde S-type planets are more dynamically stable than misaligned planets because the ZKL oscillation does not operate in a retrograde orbit, providing a more stable environment for planet formation in situ.

\begin{acknowledgements}
We kindly thank the anonymous referee for constructive comments. Computer support was provided by the DiRAC Data Intensive service at Leicester, operated by the University of Leicester IT Services, which forms part of the STFC DiRAC HPC Facility (\url{www.dirac.ac.uk}), and the Isambard 3 Tier-2 HPC Facility. Isambard 3 is hosted by the University of Bristol and operated by the GW4 Alliance (\url{https://gw4.ac.uk}), and is funded by UK Research and Innovation and the Engineering and Physical Sciences Research Council [EP/X039137/1]. This research was also enabled in part by support provided by the BC DRI Group and the Digital Research Alliance of Canada (\url{https://www.alliancecan.ca/en}; RAPI: bwj-303-ac, CCRI: kfb-843-01). We acknowledge the use of the Sarracen Python package \citep{Sarracen} for visualization in this work. CC acknowledges support from the CITA National Fellowship (grant number DIS-2022-568580). CC thanks Man Hoi Lee for the useful discussion at the CITA@40 conference. CJN acknowledges support from the Leverhulme Trust (grant number RPG-2021-380). This project has received funding from the European Union's Horizon 2020 research and innovation programme under the Marie Skłodowska-Curie grant agreement No.~823823 (Dustbusters RISE project). CC gratefully acknowledges the hospitality of Stony Brook University and the Flatiron Institute during a Dustbusters secondment. SHL acknowledges support from NASA grant 80NSSC19K0443.
\end{acknowledgements}

\bibliography{main}
\bibliographystyle{aasjournal}

\end{document}